\documentclass[12pt]{article}
\pdfoutput=1
\usepackage[top=1in, bottom=1in, left=1.in, right=1.in]{geometry}
\usepackage[english]{babel}
\usepackage{atbegshi,cite}
\usepackage{amsmath,amssymb,amsbsy,amstext, amsthm, simplewick}
\usepackage{hyperref}
\usepackage{graphicx}
\usepackage{amsfonts}
\usepackage[small]{caption}
\usepackage{upgreek}
\usepackage[titletoc]{appendix}
\usepackage{setspace}
\usepackage[dvipsnames]{xcolor}
\usepackage{slashed}

\newcommand{\newc}{\newcommand}
\newc{\fpi}{f_{\pi}}
\newc{\etap}{\eta^{\prime}}
\newc{\llll}{\langle\lambda\lambda\rangle}
\newc{\FFd}{F^a\tilde F^a}
\newc{\qbar}{{\overline q}}
\newc{\TR}{{\rm Tr}}
\newc{\Kahler}{K\"ahler }
\newc{\Zbb}{{\mathbb Z}}
\newc{\Rt}{{\mathbb R}^3}
\newc{\Rf}{{\mathbb R}^4}
\newc{\So}{{\mathbb S}^1}
\newc{\zt}{{\mathbb Z}_2}
\newc{\RtSo}{{\mathbb R}^3\times{\mathbb S}^1}
\newc{\scriminus}{{\cal I}^-}
\newc{\scriplus}{{\cal I}^+}
\newc{\mpl}{M_p}
\newc{\Ricci}{\mathcal{R}}
\newc{\bv}{\phi}
\newc{\calU}{{\cal U}}
\newc{\calK}{K}
\newc{\calUi}{{\cal U}^{-1}}
\newc{\calG}{{\cal G}}
\newc{\calM}{{\cal M}}
\newc{\calL}{{\cal L}}
\newc{\calO}{{\cal O}}
\newc{\calR}{{\cal R}}
\newc{\calQ}{{\cal Q}}
\newc{\calI}{{\cal I}}
\newc{\calOb}{{\cal O}^\dagger}
\newc{\hphi}{{\hat\phi}}
\newc{\tb}[1]{{\bf\color{MidnightBlue}TB:~#1}}
\newc{\pd}[1]{{\color{blue}#1}}

\theoremstyle{plain}
\theoremstyle{plain} 
\theoremstyle{plain} 
\theoremstyle{plain}
\theoremstyle{plain}
\theoremstyle{plain}

\renewcommand{\title}[1]{{\Large\bf\flushleft{#1}}\vspace*{3ex}\\}
\renewcommand{\author}[2]{{\noindent\hspace*{2.5em}\large#1}
                     \footnote{Electronic mail: $\mathtt{#2}$}\\}

\begin{document}
\begin{titlepage}
\begin{flushright}
{\large 
~\\
}
\end{flushright}

\vskip 2.2cm

\begin{center}

{\large \bf Path Integrals for Causal Diamonds \\ and the Covariant Entropy Principle}

\vskip 1.4cm

{{Tom Banks}$^{(a)}$,  {Patrick Draper}$^{(b)}$, and  {Szilard Farkas}}
\\
\vskip 1cm
{$^{(a)}$ Department of Physics and NHETC, Rutgers University, Piscataway, NJ 08854}\\
{$^{(b)}$ Department of Physics, University of Illinois, Urbana, IL 61801}
\vspace{0.3cm}
\vskip 4pt

\vskip 1.5cm

\begin{abstract}
We study causal diamonds in Minkowski,  Schwarzschild, (anti) de Sitter, and Schwarzschild-de Sitter spacetimes using Euclidean methods.  The null boundaries of causal diamonds are shown to map to isolated punctures in the Euclidean continuation of the parent manifold. Boundary terms around these punctures decrease the Euclidean action by $A_\diamond/4$, where $A_\diamond$ is the area of the holographic screen around the diamond. We identify these boundary contributions with the maximal entropy of gravitational degrees of freedom associated with the diamond. 
\end{abstract}

\end{center}

\vskip 1.0 cm

\end{titlepage}
\setcounter{footnote}{0} 
\setcounter{page}{1}
\setcounter{section}{0} \setcounter{subsection}{0}
\setcounter{subsubsection}{0}
\setcounter{figure}{0}

%\doublespacing
%\onehalfspacing

%\tableofcontents

\section{Introduction}

It has become increasingly clear over the past few years that the key to understanding how Einstein's theory of General Relativity fits into the framework of quantum mechanics is the relationship between quantum information and space-time geometry.  Much of the work in this area has focused on the AdS/CFT correspondence, where the Ryu-Takayanagi formula~\cite{rt} 
connects precise calculations in quantum field theory (QFT) to the areas of space-time submanifolds of the dual geometry.  The tensor network version~\cite{Swingle:2012wq,Pastawski:2015qua,Harlow:2016vwg,Swingle:2017blx,Bao:2018pvs,Bao:2019fpq} provides a way to extend these ideas to truly localized regions of space-time: causal diamonds of finite area. 

A precise relation between particular space-time geometries and hydrodynamic concepts in a quantum theory of gravity (like the entropies of subsystems)  is in tension with the idea that space-time geometry is a fluctuating quantum variable.  The entropies of large subsystems experience only small quantum fluctuations, which moreover can be completely reproduced by a {\it classical} statistical theory, like the Einstein-Smoluchowski explanation of Brownian motion, with no trace of the interference effects and associated violations of Bayes' conditional probability rule that are characteristic of quantum probabilities.
Nonetheless, evidence mounts~\cite{Saad:2019lba,Penington:2019kki,Almheiri:2019qdq} that Euclidean path integrals over geometries can reproduce features of quantum gravity that go beyond entropies of subsystems.  Thus it is fair to say that at present the relationship between the Euclidean gravitational path integral and quantum gravity remains deeply mysterious. The present paper will not resolve these mysteries, but it will extend the classes of geometric quantities that can be given a hydrodynamic interpretation.

We consider finite causal diamonds in various spacetimes and define a procedure for computing their maximal gravitational entropy $S_\diamond$ from a Euclidean action. We consider only spherically symmetric spacetimes, and in cases lacking translation invariance, we place the center of the diamond at the center of the spacetime.  In all cases our results amount to 
\begin{align}
S_\diamond=A_\diamond/4,
\label{eq:SA}
\end{align} 
where $A_{\diamond}$ is the $d -2$ volume in Planck units of the holographic screen, or the leaf of maximal $d - 2$ volume in a null foliation of the boundary of the diamond. (More precisely, we will see that Eq.~(\ref{eq:SA}) holds for ordinary diamonds in maximally symmetric spacetimes. We will also define ``cored" diamonds surrounding black holes; for these, we will find an additional contribution from the black hole horizon, which serves as an inner null boundary of the diamond.)

The Gibbons-Hawking calculation~\cite{gh} of particular black hole entropies via Euclidean path integrals can be considered evidence for the Bekenstein-Hawking area formula. Similarly, the results here can be considered evidence for the covariant entropy principle (CEP)~\cite{Jacobson:1995ab,Fischler:1998st,Bousso:1999xy,Bousso:1999cb,Bousso:1999dw}. The CEP states that {\it any} causal diamond in {\it any} $d$-dimensional Lorentzian spacetime is associated with a Hilbert space whose maximal entropy is one quarter of the area of the diamond's holographic screen:
\begin{align}
\log\,{\rm dim}\,{\cal H} = A_{\diamond}/4.
\label{eq:ceb}
\end{align}
More primitively, we view our results as an addition to the list of thus-far perplexing connections between bulk gravitational path integrals, volumes of space-time submanifolds, and precise quantum calculations.  As we will see, the prescription for computing the diamond entropies involves various choices, including, for example,  the coordinates to be analytically continued and  the signs of normal vectors on diamond boundaries. We will only motivate these choices; a true understanding of the formulae we obtain would provide a complete justification of the prescription.

We consider the semiclassical approximation to the gravitational path integral. The partition function and Euclidean action are
\begin{align}
&~~~~~~~~~~~~~~~~~~~~Z=\int Dg\, e^{-I_E[g]}\nonumber\\
I_E &= -\frac{1}{16\pi}\int_{\calM} d^4x \sqrt{g} \left(\calR-2\Lambda\right)-\frac{1}{8\pi}\int_{\partial\calM} d^3x \sqrt{h} K. 
\label{eq:action}
\end{align}
Occasionally it is useful to include another term that depends only on the boundary metric $h$, but we will not need to do so here.
 Our general strategy is as follows. First, we cover the causal diamond with inextendible ``diamond universe" coordinates. These coordinates are not particularly unique, but we will require that the time coordinate $s$ has the following properties: the holographic screen lies in constant time slice $s=0$, and $\partial_s$ is an instantaneous timelike Killing vector on that surface (this is automatic if $s=0$ is a moment of time reflection symmetry.) We then continue $s$ to Euclidean signature. In all cases we find that the  continuation is in fact the same as the Euclidean continuation of the entire parent manifold in which the diamond was embedded, with the exception of isolated punctures. These punctures are associated with the null boundaries of the diamond in Lorentzian signature, and we argue that Gibbons-Hawking-York (GHY)  terms~\cite{PhysRevLett.28.1082,gh} on infinitesimal boundaries around the punctures compute the entropies associated with the diamond horizons.

This work was motivated by the CEP and by a paper of Jacobson and Visser~\cite{JV}, who showed that a first law can be ascribed to finite causal diamonds in maximally symmetric spacetimes. We will restrict our attention to $d=4$, where the holoscreen is the maximal-area 2-surface on the boundary of the diamond, and give all results in Planck units. We will find that Euclidean techniques can be used to recover Eq.~(\ref{eq:ceb}).

Some of our discussion of boundary terms around punctures was motivated by the ADM analysis  of Euclidean Schwarzschild-de Sitter in Ref.~\cite{Teitelboim:2002cv}. In an Appendix we review this technique and show that it can be used to derive various sum rules between different boundary terms in various spacetimes. 

\section{Minkowski}

\begin{figure}[t!]
\begin{center}
\includegraphics[width=0.3\linewidth]{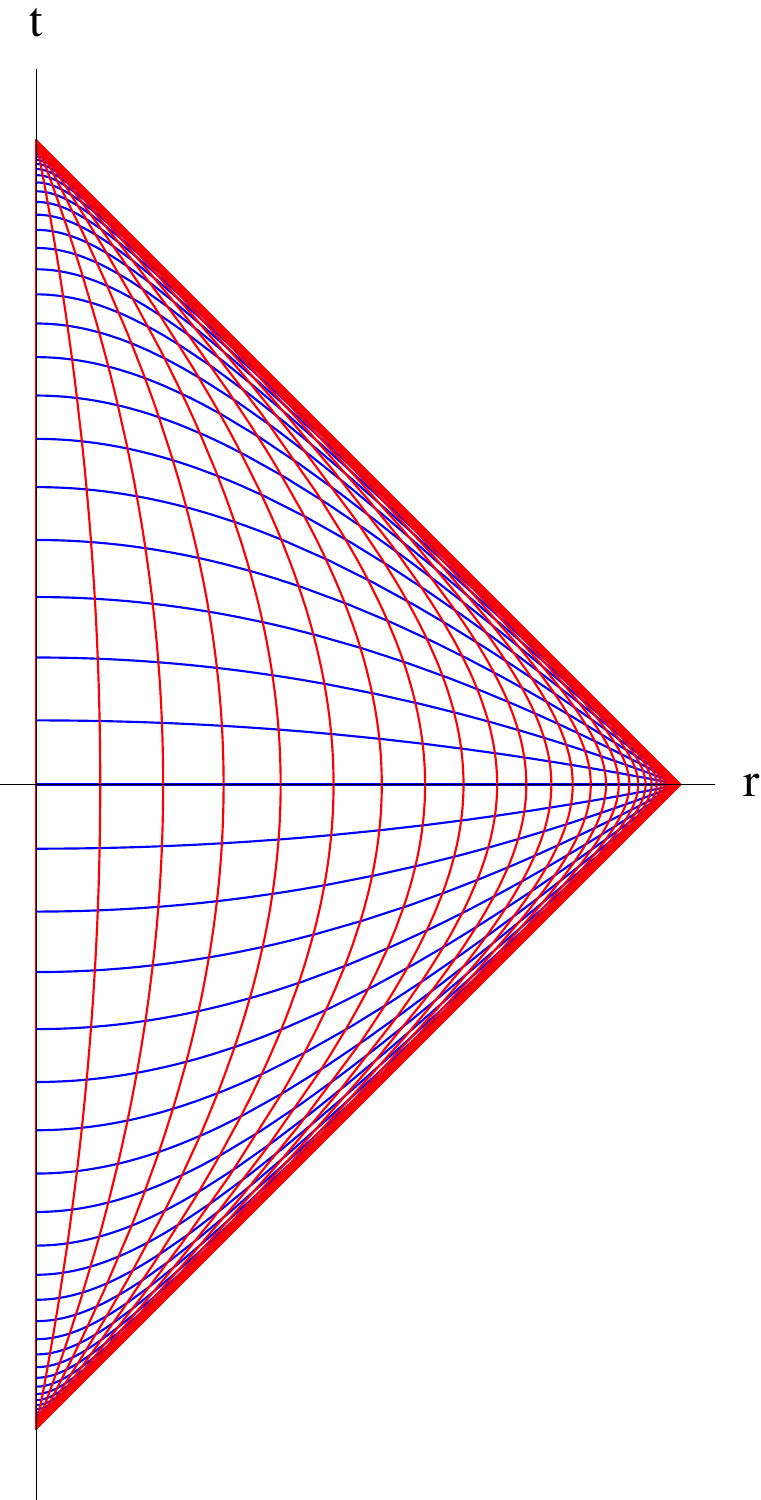}
\caption{A sketch of the diamond universe coordinates in Eq.~(\ref{eq:minktosx}). Contours of constant $x$ (red) and $s$ (blue) cover the Minkowski space diamond (an $S^2$ is suppressed). 
} 
\label{fig:minkunivcoords}
\end{center}
\end{figure}

We start with a causal diamond in  Minkowski space. We work in spherical coordinates $(t,r,\theta,\phi)$ and center a diamond of proper time $\tau$ on the origin. It is convenient to cover the diamond in inextendible coordinates introduced by Jacobson and Visser~\cite{JV},
\begin{align}
r &= \frac{\tau}{2}\left(\frac{\sinh{x}}{\cosh{x} +\cosh{s}}\right)\nonumber\\
t &=  \frac{\tau}{2}\left(\frac{\sinh{s}}{\cosh{x} +\cosh{s}}\right).
\label{eq:minktosx}
\end{align}
We will refer to these as ``diamond universe" coordinates, and we will use variations on them throughout this work. 
$s$ is a time coordinate running from $-\infty$ to $+\infty$ and $x$ is a radial coordinate running from $0$ to $\infty$. These coordinates cover the diamond, and any constant $x$ trajectory reaches the tips of the diamond at infinite $s$. A sketch is shown in Fig.~\ref{fig:minkunivcoords}. 

The line element is
\begin{align}
dl^2 &=C^2\left(-ds^2+dx^2+\sinh^2(x) d\Omega^2\right)\nonumber\\
C&=\frac{\tau/2}{\cosh(s)+\cosh(x)}.
\end{align}
Since the metric is an even function of $s$, $\partial_s$ is an instantaneous timelike Killing vector on the maximal slice of the diamond.
The Euclidean continuation $s\rightarrow -i s_E$ is
\begin{align}
dl_E^2 &= C_E^2 \left(ds_E^2+dx^2+\sinh^2(x) d\Omega^2\right)\nonumber\\
C_E&=\frac{\tau/2}{\cos(s_E)+\cosh(x)}.
\label{eq:minkcontmetric}
\end{align}
We can take $s_E$ to be periodic with period $2\pi$. 

It is instructive to compare the analytic continuation of the finite diamond to the continuation of all of Minkowski space.   What patch of $\Rf$ is covered by the continuation of the  diamond?

\begin{figure}[t!]
\begin{center}
\includegraphics[width=0.5\linewidth]{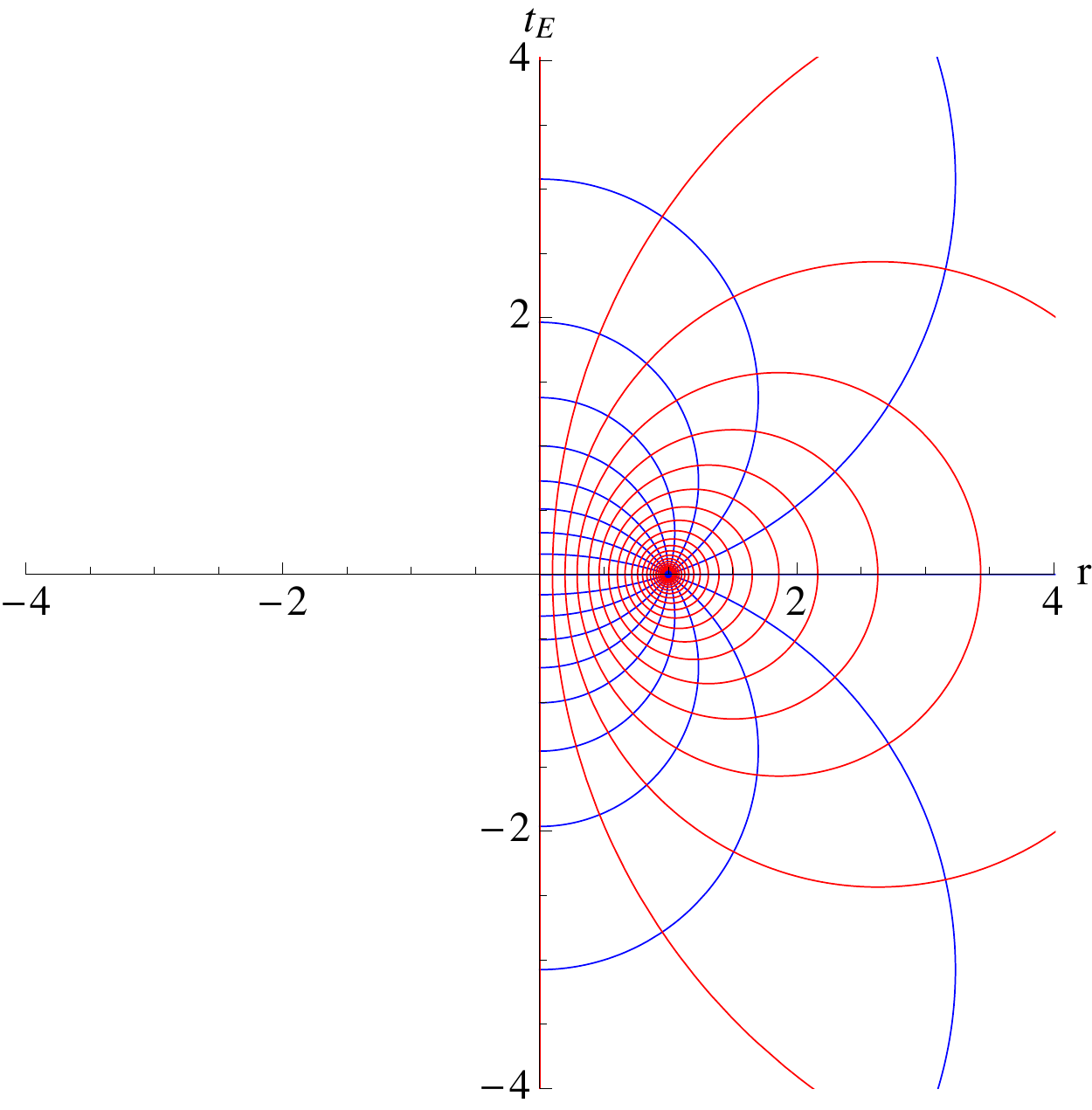}
\caption{Contours of constant $x$ (red) and $s_E$ (blue) coordinates on coordinates $(r, T_E)$ for $\Rf$ (with an $S^2$ suppressed). In this example $\tau=2$.  The continuation of the finite diamond covers almost all of $\Rf$. The boundary $x\rightarrow\infty$ corresponds to a boundary of topology $S^1\times S^2$ around $(r,t_E)=(\tau/2,0)$. The singularity at $(x,s_E)=(0,\pi)$ maps to $\infty$.
} 
\label{fig:coords}
\end{center}
\end{figure}

We define a map to $(t_E,r,\theta,\phi)$ by the continuation of Eq.~(\ref{eq:minktosx}),
\begin{align}
r &= \frac{\tau}{2}\left(\frac{\sinh x}{\cosh x+\cos s_E}\right)\nonumber\\
t_E &=  \frac{\tau}{2}\left(\frac{\sin s_E}{\cosh x+\cos s_E}\right).
\label{eq:Rfourmap}
\end{align}
Remarkably, the continuation of the finite diamond covers nearly all of $\Rf$. Only the point at $x\rightarrow\infty$ is missing, corresponding to an infinitesimal tube-like boundary of topology $S^1\times S^2$ around $(r,t_E)=(\tau/2,0)$.  The map~(\ref{eq:Rfourmap}) is sketched in Fig.~\ref{fig:coords}.

The metric~(\ref{eq:minkcontmetric}) also has a singularity at  $(x,s_E)=(0,\pi)$. Eq.~(\ref{eq:Rfourmap}) maps this singularity to $r=\infty$ in $\Rf$:  the singularity is in fact equivalent to the ordinary large-radius boundary in the continuation of all of Minkowski space.

Thus, the only difference between the continuation of the finite diamond and the continuation of Minkowski spacetime is the presence of an  infinitesimal boundary of topology $S^1\times S^2$ around $(r,t_E)=(\tau/2,0)$. This point  corresponds to $x\rightarrow\infty$ in diamond universe coordinates, or the horizon of the diamond in Lorentzian signature. It is therefore natural to identify the Euclidean action of the diamond  with a GHY term on this boundary.

This identification may appear unusual and it is worth dwelling on it for a moment. Ordinarily, the thermodynamics properties of horizons are not computed in this way, by the insertion of a boundary term at the Euclidean continuation of the horizon. For example, in Euclidean Schwarzschild the free energy is usually computed from a boundary term at infinity, rather than at  $r=2M$. In Euclidean de Sitter, the entropy is computed from the bulk Einstein-Hilbert term, rather than a boundary term at the cosmological horizon $r=L$. However, both the black hole and cosmological horizon entropies can also be computed from the GHY term on infinitesimal boundaries at $r=2M$ and $r=L$, respectively, with outward normal pointing toward the bulk. This result can be derived from the ADM formulation of the action~\cite{Banados:1993qp,Teitelboim:2002cv}, which we review in the Appendix. In this light, our prescription to compute the entropy of the diamond horizon from a boundary term at the continuation of the horizon is not so unprecedented.

In which direction is the boundary surface oriented?  The map to $\Rf$  compels us to orient the surface  \emph{outward} from the puncture at $(r,t_E)=(\tau/2,0)$, so that its orientation matches the orientation of the usual boundary at infinity. This also consistent with the prescription for computing black hole and cosmological horizon entropies described in the previous paragraph. More covariantly, we compute the full gravitational action in a small neighborhood in $\Rf$ around the point $(r,t_E)=(\tau/2,0)$: the Einstein-Hilbert term vanishes in the infinitesimal limit, leaving the outward-directed GHY term.  In diamond universe coordinates, this corresponds to an \emph{inward} pointing boundary at $x\rightarrow\infty$, pointing toward \emph{smaller} $x$.

The boundary term is easy to compute in either static or diamond universe coordinates; we will use the latter. 
%The trace of the extrinsic curvature and metric (g) and induced metric (h) determinants are 
%\begin{align}
%K&=n^a_{;a}=-\frac{1}{\sqrt{g}}\partial_x\left( \sqrt{g}C_E^{-1}\right)|_{x\rightarrow\infty}\nonumber\\
%\sqrt{g}& = C_E^4\sinh^2(x)\sin(\theta)\nonumber\\
%\sqrt{h}& = C_E^3\sinh^2(x)\sin(\theta).
%\end{align}
We obtain
\begin{align}
\sqrt{h}K =- \tau^2\sin(\theta)\sinh(x)\left(\frac{2\cosh(x)(\cos(s)+\cosh(x))-3\sinh^2(x)}{4(\cos(s)+\cosh(x))^3}\right),
\end{align}
and the limiting behavior at large $x$ is
\begin{align}
\sqrt{h}K\rightarrow \frac{1}{4}\tau^2\sin(\theta)~~~{\rm~as~}x\rightarrow\infty.
\end{align}
Thus the boundary action is
\begin{align}
I_{GHY} &=- \left(\frac{1}{4}\tau^2\right)\frac{1}{8\pi}\int_0^{2\pi}ds_E \int d\Omega\nonumber\\
&=-\frac{1}{4}\pi\tau^2\nonumber\\
&=-A_\diamond/4
\label{eq:minklargex}
\end{align}
where $A_\diamond=4\pi(\tau/2)^2$ is the area of the holographic screen, the boundary of the maximal slice.

We interpret this result as the  free energy $\beta_\diamond F_\diamond = \beta_\diamond E_\diamond - S_\diamond= - S_\diamond$ of the quantum gravitational degrees of freedom in the causal diamond:\footnote{These are not to be confused with quantum field theoretic degrees of freedom associated with the gravitational field.  They are the as-yet-undetermined underlying variables of a general quantum theory of gravity.  We are uncovering only their hydrodynamic properties.}
\begin{align}
S_\diamond=-I_{Euclidean} = A_\diamond/4,
\end{align}
consistent with the CEP.   The identification of the free energy and the entropy is a consequence of the fact that at $x = \infty$ the radius of the thermal circle is zero: the temperature is infinite, and the partition function simply counts all of the states in the Hilbert space.\footnote{ To interpret the GHY term as the free energy of an equilibrium thermodynamic system, we have to specify  coordinates for which the time coordinate is periodic on the boundary around the puncture. This requirement is satisfied by the $x, s_E$ coordinates, and the temperature is infinite because the proper periodicity is zero at the puncture.}  This is reminiscent of the derivation of the CEP following Jacobson~\cite{Jacobson:1995ab}, where the entropy and energy that are used to relate Einstein's equations and the first law of thermodynamics are those appropriate to an infinite temperature Unruh trajectory.

\section{de Sitter}
\label{sec:dS}
Next we consider causal diamonds in the other maximally symmetric spacetimes, beginning with de Sitter. The static patch metric is 
\begin{align}
dl^2 &= -f(r) dt^2 +f(r)^{-1} dr^2 + r^2 d\Omega^2\nonumber\\
f(r) &= 1-\left(\frac{r}{L}\right)^2
\end{align}
and we introduce a radial tortoise coordinate
\begin{align}
r=L\tanh(r_*/L).
\end{align}
We denote the tortoise coordinate radius of the holographic screen as $R_*$. Its area radius is $R=L\tanh(R_*/L)$ and the area is $A_\diamond=4\pi R^2$. 

For variety, we will use a different set of diamond universe coordinates in this case. These coordinates were also introduced by Jacobson and Visser in the study of conformal Killing vectors preserving causal diamonds~\cite{JV}.  However, this property is inessential; in particular, there is no conformal Killing vector preserving the causal diamond studied in the Schwarzschild case below, and we will use a different set of coordinates in the AdS case.

The transformation from the finite diamond to the $s,x$ variables defined in~\cite{JV} is given in terms of lightcone coordinates $u = t-r_*$, $v=t+r_*$, $\bar u = s-x$, $\bar v = s+x$:
\begin{align}
e^{u/L} &= \frac{\cosh[(R_*/L+\bar u)/2]}{\cosh[(R_*/L-\bar u)/2]}\nonumber\\
e^{v/L} &= \frac{\cosh[(R_*/L+\bar v)/2]}{\cosh[(R_*/L-\bar v)/2]}.
\end{align}
Here $x$ runs from zero to infinity. 

The metric in $s,x$ coordinates is
\begin{align}
dl^2 &=C^2\left(-ds^2+dx^2+\sinh^2(x) d\Omega^2\right)\nonumber\\
C &= \frac{L\sinh(R_*/L)}{\cosh(s)+\cosh(x)\cosh(R_*/L)}.
\end{align}
It can be checked that in the limit $R_*\rightarrow\infty$, making the substitution $r=L\tanh x$ recovers the ordinary dS static patch with time $s$. 

The Euclidean continuation $s\rightarrow -i s_E$ is
\begin{align}
dl_E^2 &= C_E^2 \left(ds_E^2+dx^2+\sinh^2(x) d\Omega^2\right)\nonumber\\
C_E&=\frac{L\sinh(R_*/L)}{\cos(s_E)+\cosh(x)\cosh(R_*/L)}.
\end{align}
Again we can take $s_E$ to be periodic with period $2\pi$. The Euclidean continuation of the original time coordinate, $t\rightarrow -i t_E$, is related to $x, s_E$ by
\begin{align}
\tan(t_E/L) = \frac{\sin(s_E) \sinh (R_*/L)}{\cos(s_E) \cosh (R_*/L)+\cosh(x)}.
\end{align}
We can make the right-hand side arbitrarily small by tuning $s_E$ close to zero, or arbitrarily large by taking $x$ larger than $R_*/L$ and tuning $\cos(s_E)$ close to $\cosh(x)/\cosh(R_*/L)$. Therefore, $t_E/L$ is an ordinary angle. This is also what is found in the usual analysis of Euclidean dS, requiring the absence of a conical singularity at $r=L$. The tortoise coordinate satisfies
\begin{align}
r_* = \frac{L}{2}\log\left(\frac{\cos(s_E)+\cosh(x+R_*/L)}{\cos(s_E)+\cosh(x-R_*/L)}\right).
\end{align}
For any $s_E$, we can make $r_*$ arbitrarily negative or positive by tuning $x$.

Therefore, as in the Minkowski case, the $x, s_E$ coordinates cover the entire Euclidean dS manifold with the exception of the surface at $x\rightarrow \infty$, which is the null boundary of the diamond in Lorentzian signature. In  Euclidean static patch tortoise coordinates, this boundary maps to the point $t_E=0$, $r_*=R_*$.

We associate the puncture with the null boundary of the diamond, and to compute the free energy we place an infinitesimal \emph{outward}-pointing boundary around the puncture (corresponding in the diamond universe coordinates to a boundary at $x\rightarrow\infty$ pointing toward smaller $x$). The limiting behavior of the associated GHY term is
\begin{align}
\sqrt{h}K\rightarrow R^2\sin(\theta)~~~{\rm~as~}x\rightarrow\infty,
\end{align}
so it contributes 
\begin{align}
I_{GHY}=  -A_\diamond /4.
\end{align}
Again  we interpret this result as $\log(Z)$ at infinite temperature, counting the quantum gravitational degrees of freedom associated with a finite causal diamond, $S_\diamond= A_\diamond/4$.

dS has a maximal causal diamond, $R_*\rightarrow \infty$. In this limit the diamond radius in ordinary static patch coordinates is $r=L$, filling the spacetime. We see that we recover the entropy of the cosmological horizon in this limit, $S_{dS}=A_{dS}/4$.

Ordinarily, the dS horizon entropy is computed from a Euclidean bulk Einstein-Hilbert term, and it is said that there are no boundaries, since Euclidean dS is topologically $S^4$. Here we obtained the dS entropy purely from a boundary term around the ``point" at $r=L$. This suggests an interesting bulk-boundary sum rule. As mentioned above, this is not an accident, and is one of a family of sum rules that can be derived by equating the ADM calculation of the action to the Einstein-Hilbert calculation. We derive this relationship in the Appendix.

\section{Anti de Sitter}
We compute the maximal entropy of the AdS diamond using diamond universe coordinates similar to those in the Minkowski case, which for AdS are not adapted to the timelike conformal Killing vector.  Indeed, any coordinates that agree with the ones we use near the boundaries of the diamond would work just as well. In AdS we start in global coordinates, for which the Euclidean AdS metric is
\begin{align}
dl^2 = \frac{1}{\cos^2(\rho/L)}\left(dt_E^2+d\rho^2+L^2\sin^2(\rho/L)d\Omega^2\right)
\end{align}
where $0\leq \rho \leq \pi L/2$. For a causal diamond with coordinate radius $\rho=\rho_d$ on the holographic screen, we define the diamond universe coordinates in a manner similar to the flat space expressions in Eq.~(\ref{eq:minktosx}), substituting $r\rightarrow\rho$, $\tau\rightarrow 2\rho_d$. The analytically continued diamond and global coordinates are related by
\begin{align}
\rho &= \rho_d\left(\frac{\sinh x}{\cosh x+\cos s_E}\right)\nonumber\\
\tau_E &=  \rho_d\left(\frac{\sin s_E}{\cosh x+\cos s_E}\right).
\label{eq:AdSmap}
\end{align}
$s_E$ is bounded for small $x$ by $\cos s_E > -\cosh x+\frac{2\rho_d}{\pi L}\sinh x$, which maps to the boundary of Euclidean AdS at  $\rho=\pi L/2$.
Again, the continuation of the finite diamond covers nearly all of Euclidean AdS. Only the surface at $x\rightarrow\infty$ is missing, corresponding to an infinitesimal tube around $(\rho,\tau_E)=(\rho_d,0)$.

As before, we compute the GHY term from an inward-pointing boundary at $x\rightarrow\infty$. The limiting behavior is
\begin{align}
&\sqrt{h}K\rightarrow L^2\tan^2(\rho_d/L)\sin(\theta)~~~{\rm~as~}x\rightarrow\infty.
\end{align}
$L\tan (\rho_d/L)$ is the area radius of the holographic screen (the radius $r$ defined by setting the proper area equal to $4\pi r^2$), so we find that this new boundary has Euclidean action
\begin{align}
I_{GHY}  = -A_\diamond/4,
\end{align}
with a similar interpretation to the previous cases.

\section{Schwarzschild}

Now we consider cases where there is a black hole at the center of the spatial slices in the diamond. We will not include the interior of the black hole in the diamond, and in this sense the diamond is ``cored." This introduces a new complication,  a second null boundary in Lorentzian signature, lying on the black hole horizon. We will see that both horizons are mapped to punctures under Euclidean continuation of the inextensible coordinates. 

\begin{figure}[t!]
\begin{center}
\includegraphics[width=0.45\linewidth]{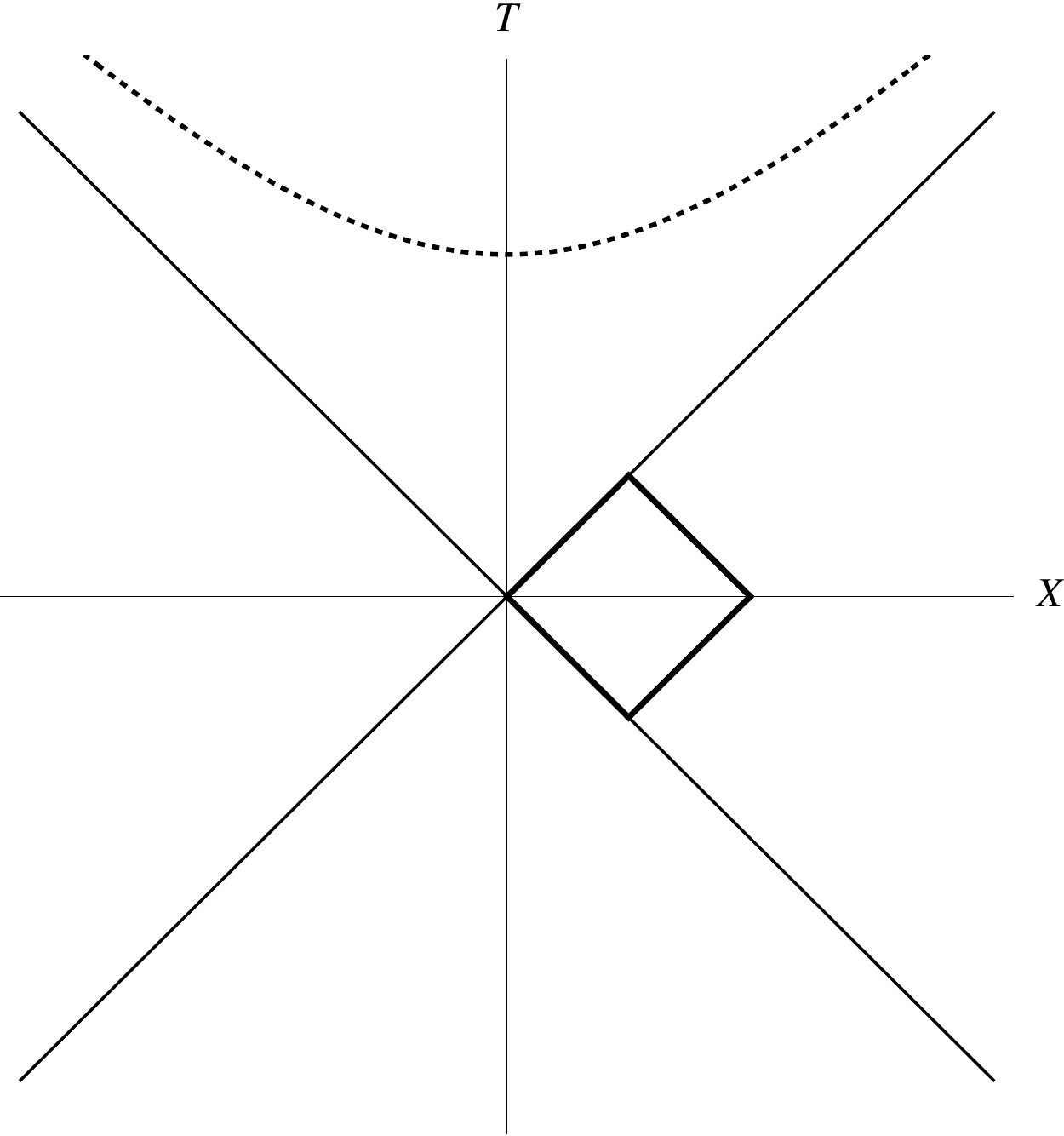}~~~~~~~~
\includegraphics[width=0.48\linewidth]{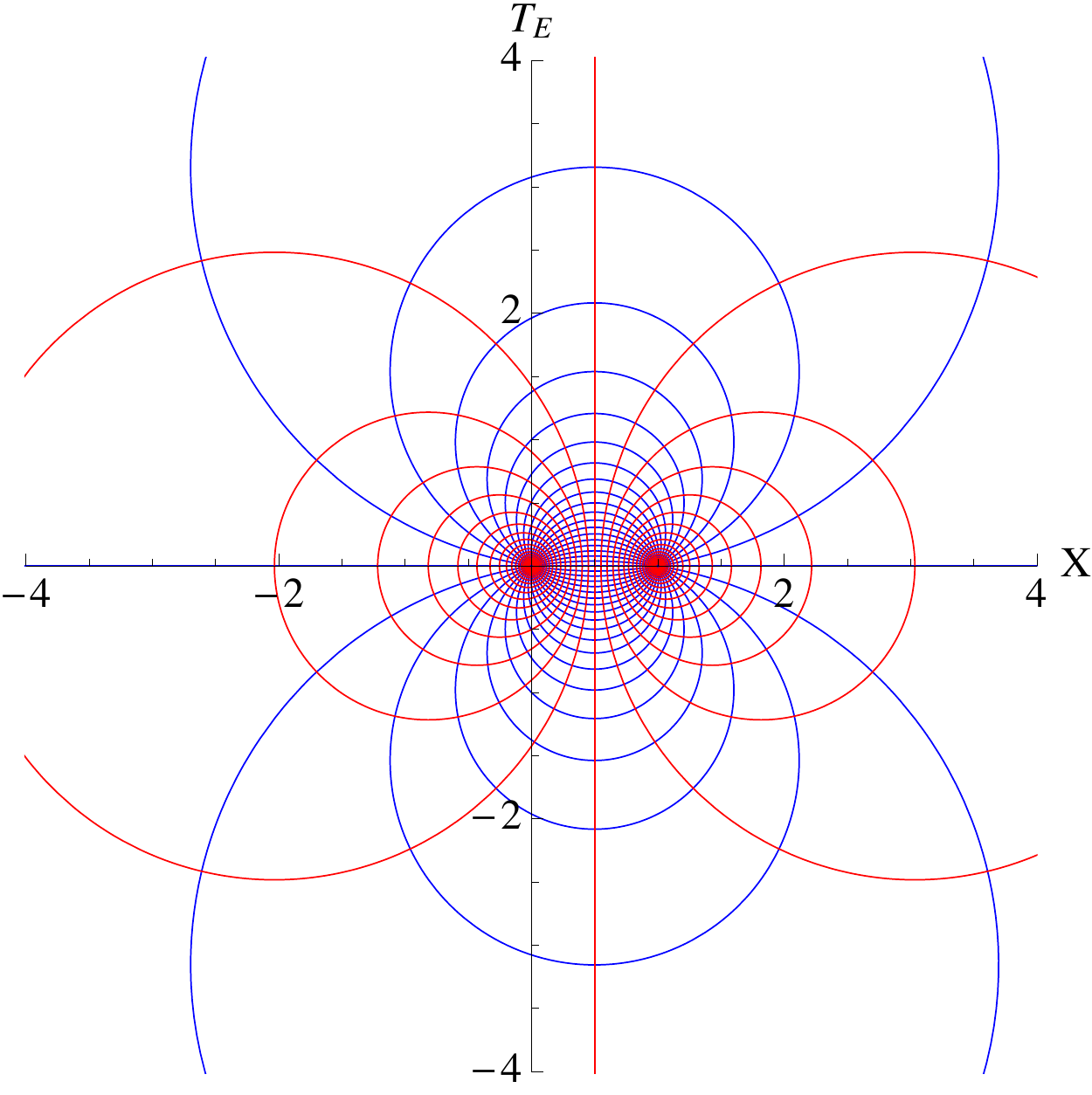}
\caption{Left: Causal diamond surrounding a Schwarzschild black hole in Kruskal-Szekeres coordinates (with an $S^2$ suppressed). Right: Contours of constant $x$ (red) and $s_E$ (blue) coordinates, which define the Euclidean continuation of the finite diamond around the black hole, on Euclidean Kruskal-Szekeres coordinates. In this example $\tau=2$. The Euclideanized diamond covers almost all of Euclidean Schwarzschild, except for the points $(X,T_E)=(\tau/2,0)$ and $(0,0)$, corresponding to $x=\pm\infty$.  The singularity at $(x,s_E)=(0,\pi)$ maps to $r=\infty$, where $r$ is the usual Schwarzschild coordinate, a function of $X^2+T_E^2$.
} 
\label{fig:schwcoords}
\end{center}
\end{figure}

First we consider a diamond surrounding a Schwarzschild black hole.
It is convenient to start from the Kruskal-Szekeres coordinates,
\begin{align}
&dl^2 = \frac{32M^3}{r}e^{-r/2M}(-dT^2+dX^2)+r^2d\Omega^2\nonumber\\
&T^2-X^2=(1-r/2M)e^{r/2M}.
\end{align}
The second equation gives an implicit definition of $r$. 
The region outside the black hole is $X>0$, $T^2-X^2<0$. Therefore, while a causal diamond in flat space looks like half a diamond on a (radius, time) plot (because $r>0$),  a causal diamond with a black hole at the center, in Kruskal-Szekeres coordinates and with coordinate radius $\Delta X=\tau/2$ on the maximal slice,  looks like a full diamond  offset to the right  on a plot in  $(X,T)$ coordinates. This is shown in the left panel of Fig.~\ref{fig:schwcoords}.

The  radius of the diamond $r_{\tau/2}$ satisfies the equation above with $X=\tau/2$ and $T=0$, i.e. 
\begin{align}
-\tau^2/4=(1-r_{\tau/2}/2M)e^{r_{\tau/2}/2M}.
\end{align}
 We can  introduce the diamond universe coordinates in a manner similar to the previous case,
\begin{align}
X-\frac{\tau}{4} &=  \frac{\tau}{4}\left(\frac{\sinh{x}}{\cosh{x} +\cosh{s}}\right)\nonumber\\
T &= \frac{\tau}{4}\left(\frac{\sinh{s}}{\cosh{x} +\cosh{s}}\right).
\end{align}
$s$ still runs from $-\infty$ to $+\infty$, but $x$ now runs from $-\infty$ to $+\infty$ as well.
In these coordinates,
\begin{align}
&dl^2 = C^2(-ds^2+dx^2)+r^2d\Omega^2\nonumber\\
&C=\left(\frac{\tau/4}{\cosh(s)+\cosh(x)}\right)\left(\frac{32M^3}{r}e^{-r/2M}\right)^{1/2}\nonumber\\
-&\frac{\tau ^2 e^x}{8 (\cosh (s)+\cosh (x))}=(1-r/2M)e^{r/2M}.
\end{align}
Since the metric is an even function of $s$, $\partial_s$ is again an instantaneous timelike Killing vector on the maximal slice.
The Euclidean continuation $s\rightarrow -i s_E$ is
\begin{align}
&dl^2 = C_E^2(ds_E^2+dx^2)+r^2d\Omega^2\nonumber\\
&C_E=\left(\frac{\tau/4}{\cos(s_E)+\cosh(x)}\right)\left(\frac{32M^3}{r}e^{-r/2M}\right)^{1/2}\nonumber\\
-&\frac{\tau ^2 e^x}{8 (\cos (s_E)+\cosh (x))}=(1-r/2M)e^{r/2M}.
\end{align}
Again we can take $s_E$ to be periodic with period $2\pi$. 

We can also define Euclidean Kruskal-Szekeres coordinates by the continuation $T\rightarrow -i T_E$. This continuation specifies the periodicity of the Euclidean Schwarzschild coordinate time $t_E$: $\tan(t_E/4M) = T_E/X$, so if we fix $T_E^2+X^2$, we see that $t_E/4M$ is an ordinary angle. This produces the correct periodicity to avoid a conical singularity at $r=2M$. 

The natural map between the Euclidean diamond coordinates and the Euclidean Kruskal-Szekeres coordinates is
\begin{align}
X&=\frac{\tau}{4}\left(1+\frac{\sinh x}{\cos s_E+\cosh x}\right)\nonumber\\
T_E&=\frac{\tau}{4}\left(\frac{\sin s_E}{\cos s_E+\cosh x}\right).
\end{align}
As in the previous cases, the Euclidean diamond corresponds to almost all of Euclidean Schwarzschild.  The surfaces at $x\rightarrow\pm \infty$ correspond to infinitesimal tubes of topology $S^1\times S^2$ around $(X,T_E)=(\tau/2,0)$ and $(0,0)$, respectively. Fig.~\ref{fig:schwcoords} shows an example of the map. The map is again singular at $(x,s_E)=(0,\pi)$, which maps to infinity in the $(X,T_E)$ plane, corresponding to the asymptotic boundary of Euclidean Schwarzschild.

We  place GHY terms around the  points at $x\rightarrow\pm \infty$ and interpret them as  the free energy seen by an observer skirting the diamond horizon.
The $x\rightarrow\infty$ boundary $(X,T_E)=(\tau/2,0)$ is computed as before: we interpret the point as representing the outer null boundary of the diamond, and choose the orientation of the boundary so that it points toward radial infinity in Kruskal-Szekeres coordinates. This corresponds to an outward-pointing boundary around  $(X,T_E)=(\tau/2,0)$. 
We have
\begin{align}
\sqrt{h}K&=-\frac{\sqrt{h}}{\sqrt{g}}\partial_x\left( \sqrt{g}C_E^{-1}\right)|_{x}\nonumber\\
%\sqrt{g}& = C_E^2 r^2\sin(\theta)\nonumber\\
%\sqrt{h}& = C_E r^2\sin(\theta)
%\end{align}
%and we obtain
%\begin{align}
%\sqrt{h}K 
&= -\frac{\sin (\theta ) \left(M \tau ^2 e^{-\frac{r}{2 M}} (6 M-r) \left(e^x \cos
   (s_E)+1\right)-8 r^2 \sinh (x) (\cos (s_E)+\cosh (x))\right)}{8 (\cos
   (s_E)+\cosh (x))^2}\nonumber\\
%\end{align}
%The limiting behavior is
%\begin{align}
%\sqrt{h}K
&\rightarrow r_{\tau/2}^2\sin(\theta)~~~{\rm~as~}x\rightarrow\infty.
\label{eq:schwdiamondarea}
\end{align}

Now we consider the boundary at $x\rightarrow  -\infty$. This corresponds to the null black hole horizon, so  this horizon maps to $(X, T_E)=(0,0)$ under Euclidean continuation. We draw an infinitesimal disc around the point, compute the outward-pointing (positive $x$-directed) GHY term, and add it to~(\ref{eq:schwdiamondarea}). We can reuse the above computation of $\sqrt{h}K$, and the limiting behavior is
\begin{align}
\sqrt{h}K\rightarrow (2M)^2\sin(\theta)~~~{\rm~as~}x\rightarrow-\infty.
\end{align}

Putting the two boundary terms together, we obtain  the  action
\begin{align}
 I_{GHY} &=-\frac{1}{8\pi}\int_0^{2\pi} d\phi\int_0^{\pi}d\theta\int_0^{2\pi} ds_E \, \left((r_{\tau/2}^2+4M^2)\sin \theta\right)\nonumber\\
&=-\pi \left(r_{\tau/2}^2+4M^2\right)\nonumber\\
&=-(A_\diamond+A_{BH})/4.
\label{eq:deltaIschw}
\end{align}
The entropy  is $(A_\diamond+A_{BH})/4$. We interpret this result as counting the total maximal entropy associated with both the inner and outer diamond horizons. This will also reproduce the expected result for the maximal diamond  in the Schwarzschild-de Sitter case considered below.

In a sense there is a term ``missing" from the above calculation, which one might have expected from experience with black holes in dS space.  Insertion of a black hole into empty dS space causes the cosmological horizon area to shrink by an amount larger than the black hole entropy, so that the total entropy decreases.  In the calculation above, we have used the fact that Minkowski space with or without a black hole has causal diamonds of arbitrarily large area,  computing the entropy of a cored diamond of fixed area.  Verlinde~\cite{Verlinde:2016toy} has shown how to obtain the shrinkage of area of a Minkowski diamond when a black hole is inserted in it from a direct Lorentzian computation. We have not attempted a Euclidean version of his calculation.

\section{Schwarzschild-de Sitter}
We conclude with the Schwarzschild-de Sitter spacetime. SdS is a thermodynamically interesting system, exhibiting two temperatures $T_b$, $T_c$ and two entropies $S_b$, $S_c$ associated with the black hole and cosmological horizons, respectively. It has been argued that SdS should be thought of as a constrained state of the empty dS ensemble~\cite{Banks:2006rx,Banks:2010tj,Banks:2013fr}. This interpretation is supported by the behavior of the total entropy, $S_{tot}=S_b+S_c\sim S_{dS}-M/T_{dS}$ for small $M$ and fixed cosmological constant, and by various other thermodynamic properties~\cite{Johnson:2019vqf,Dinsmore:2019elr,Johnson:2019ayc,Qiu:2019qgp}. The  non-equilibrium nature of SdS is reflected in the Euclidean continuation by the fact that we can  remove the conical singularity at the black hole horizon $r=r_b$, or at the cosmological horizon   $r=r_c$, but not both. The other point must be omitted, introducing an additional boundary~\cite{Teitelboim:2002cv}. The choice is fixed by taking $\beta = 1/T_b$ or $1/T_c$, corresponding to the thermodynamic ensembles associated with the black hole horizon and the cosmological horizon, respectively.

It will turn out that for the Euclidean analysis of causal diamonds in SdS, we will not need to specify $\beta$, but we will recover the total entropy 
$S_b+S_c$ for the maximal diamond. This simplification is related to the divergence of the Unruh temperatures on trajectories around the diamond boundaries.

The SdS static patch metric is
\begin{align}
dl^2&= -f(r) dt^2 + f(r)^{-1} dr^2+r^2 d\Omega^2\nonumber\\
f(r)&=1-2M/r-(r/L)^2.
\end{align}
The tortoise coordinate satisfies
\begin{align}
dr_* = \frac{dr}{f(r)}
\end{align}
or
\begin{align}
r_* = \frac{L^2}{3} \left(\frac{r_b\log(r-r_b)}{L^2/3 - r_b^2}+\frac{r_c\log(r-r_c)}{L^2/3 - r_c^2}+\frac{r_n\log(r-r_n)}{L^2/3 - r_n^2}\right).
\end{align}
Here $r_{b,c}$ are the black hole and cosmological horizon radii and $r_n = -r_b-r_c$. These radii are related to the mass and de Sitter radius as
\begin{align}
m&=\frac{r_b r_c(r_b+r_c)}{2(r_b^2+r_br_c+ r_c^2)}\nonumber\\
L^2&=r_b^2+r_br_c+r_c^2.
\end{align}
In the tortoise coordinate, the metric is 
\begin{align}
dl^2=f(r) (-dt^2+dr_*^2)+r^2d\Omega^2
\end{align}
with $r$ and $r_*$ related as above. Now we change to Kruskal-type coordinates $T,X$. In terms of the lightcone tortoise coordinates $u=t-r_*$, $v=t+r_*$ we introduce 
\begin{align}
U=-e^{-2\pi u/\beta},\;\;\;\;
V=e^{2\pi v/\beta},\;\;\;\;
T=\frac{U+V}{2},\;\;\;\;
X=\frac{V-U}{2}.\;\;\;\;
\end{align}
$\beta$ is a parameter related to the periodicity of the Euclidean continuation of the original time coordinate, $t_E$. 
We have
\begin{align}
T^2-X^2=UV=-e^{4\pi r_*/\beta}
\end{align}
and the metric is
\begin{align}
dl^2=\frac{\beta^2}{4\pi^2} f(r)e^{-4\pi r_*/\beta}(-dT^2+dX^2)+r^2d\Omega^2.
\end{align}
The future horizon is located along $T=X$, and the exterior of the black hole is $X>0$, $T^2-X^2<0$. 

Now we consider a causal diamond centered on the black hole, as in the Schwarzschild case, with coordinate radius $\Delta X=\tau/2$ on the maximal slice.  We introduce the same diamond coordinates,
\begin{align}
X-\frac{\tau}{4} &=  \frac{\tau}{4}\left(\frac{\sinh{x}}{\cosh{x} +\cosh{s}}\right)\nonumber\\
T &= \frac{\tau}{4}\left(\frac{\sinh{s}}{\cosh{x} +\cosh{s}}\right).
\end{align}
$s$  and $x$ run from $-\infty$ to $+\infty$. 
The Euclidean continuation $s\rightarrow -i s_E$ yields the metric
\begin{align}
&dl^2 = C_E^2(ds_E^2+dx^2)+r^2d\Omega^2\nonumber\\
&C_E=\left(\frac{\tau/4}{\cos(s_E)+\cosh(x)}\right)\left(\frac{\beta^2}{4\pi^2}f(r)e^{-4\pi r_*/\beta}\right)^{1/2}\nonumber\\
&e^{4\pi r_*/\beta}=\frac{\tau ^2 e^x}{8 (\cos (s_E)+\cosh (x))}.
\end{align}
 $s_E$ is periodic with period $2\pi$. Since $T_E/X = \tan(2\pi t_E/\beta) \rightarrow \tan(s_E) $ at $r_b$, and $s_E$ is an ordinary angle, we have $t_E\sim t_E+\beta$. 

As in the asymptotically flat Schwarzschild case, the Euclidean diamond coordinates map to Euclidean K-S coordinates via
\begin{align}
X-\frac{\tau}{4}&=\frac{\tau}{4}\left(\frac{\sinh x}{\cos s_E+\cosh x}\right)\nonumber\\
T_E&=\frac{\tau}{4}\left(\frac{\sin s_E}{\cos s_E+\cosh x}\right).
\end{align}
The continued diamond again covers all of Euclidean SdS, apart $x\rightarrow \infty$, corresponding to  $(X,T_E)=(\tau/2,0)$, and  $x\rightarrow -\infty$, corresponding to $(X,T_E)=(0,0)$. We identify these points with the diamond horizons and compute their entropies from GHY terms on infinitesimal boundaries around the points, choosing the orientations as in the Schwarzschild case. For the first boundary, the normal points toward smaller $x$, corresponding to an outward normal around the point $(\tau/2,0)$ in Euclidean K-S coordinates, in the same direction as the ordinary boundary at radial infinity. The second boundary  also have an outward pointing normal around $(0,0)$ in Euclidean K-S coordinates.

The GHY integrand is
\begin{align}
\sqrt{h}K =-
\frac{r \sin \theta  \left(\frac{\tau ^2}{\pi} e^{-\frac{4\pi r_*}{\beta}} \left(e^x
   \cos s_E+1\right) \left(4 \beta
   f +\beta r f'-4\pi r\right)-64 r \sinh x (\cos s_E+\cosh x)\right)}{64
   (\cos s_E+\cosh x)^2}
\end{align}
with limiting behavior
\begin{align}
\sqrt{h}K&\rightarrow r_{\tau/2}^2\sin \theta~~~{\rm~as~}x\rightarrow\infty\nonumber\\
\sqrt{h}K&\rightarrow r_{b}^2\sin \theta~~~{\rm~as~}x\rightarrow-\infty
\end{align}
Thus we obtain
\begin{align}
I_\diamond  = -(A_\diamond+A_b)/4,
\label{eq:IdiamondSdS}
\end{align}
or a  total entropy equal to $1/4$ the sum of the outer diamond horizon area and the black hole horizon area.
We see that we did not actually have to specify $\beta$ to derive the maximal diamond entropy. As $x\rightarrow\pm\infty$, the proper size of the $s_E$ thermal circle goes to zero, corresponding to an infinite temperature observer.

Let us compare Eq.~(\ref{eq:IdiamondSdS}) to the total entropy of SdS, reviewing the Euclidean computation of the latter. We can start with the ensemble associated with the cosmological horizon, setting $\beta=1/T_c$.  The   Euclidean action  receives a bulk Einstein-Hilbert contribution and a boundary contribution from $r\rightarrow r_b$, where a point is deleted to remove a conical singularity. 
The bulk term is
\begin{align}
(\Delta I_E)_{bulk} &= -\frac{1}{16\pi}\int \sqrt{g} (\calR-2\Lambda)\nonumber\\
&= \frac{1}{T_c} \left(\frac{r_b^3-r_c^3}{2L^2}\right).
\end{align}
The horizon temperatures are $T=|f'(r)|/4\pi$, or
\begin{align}
T_b=\frac{(r_c-r_b)(2r_b+r_c)}{4\pi L^2 r_b},~~~~~T_c=\frac{(r_c-r_b)(2r_c+r_b)}{4\pi L^2 r_c}.
\end{align}
The boundary term is
\begin{align}
&\sqrt{h}K = \left(3r^3/L^2 +3M-2r\right)\sin\theta,\nonumber\\
&(\Delta I_E)_{r_b} = \pi\frac{r_b r_c (2r_b+r_c)}{r_b+2r_c}
\end{align}
using the expressions for $L$ and $M$ as a function of the horizon radii. So the total is
\begin{align}
(\Delta I_E)_{bulk}+(\Delta I_E)_{r_b} = -\pi r_c^2,
\end{align}
precisely the expectation for the cosmological horizon entropy, $\beta F = -S_{c}=-A_c/4$. The puncture at $r_b$ has removed the contribution of the black hole. A similar analysis with $\beta=1/T_b$ shows that the black hole has the usual entropy $-\pi r_b^2$. So the total entropy is
\begin{align}
S_{tot} =  (A_c+A_b)/4.
\label{eq:maxsds}
\end{align}

Thus, as  the diamond radius approaches the cosmological horizon (the maximal causal diamond limit) the maximal diamond entropy $(A_\diamond+A_b)/4$ converges to the total SdS entropy in Eq.~(\ref{eq:maxsds}). 

\section{Discussion}
We have demonstrated that the Covariant Entropy Principle, relating the number of degrees of freedom associated with a finite causal diamond to the area of its holographic screen, can be obtained from a Euclidean path integral over the gravitational field.  The interesting technical feature is that the null boundaries of diamonds are mapped to punctures in the Euclideanization of the spacetime in which the diamond is embedded. The maximal entropy is associated with boundary terms around the punctures.

This result joins a long list of computations in which the ``low energy effective field theory" seems to know more about the microstates of quantum gravity than one might have expected.  A derivation of the CEP from effective field theory is particularly surprising because any quantum field theory calculation of the entropy of a diamond by summing over states gives an infinite answer.   The more general puzzle, of which this is only an example, is that the relations we derive from such Euclidean computations relate particular space-time geometries to coarse grained properties of microstates: the entropies of large subsystems.  In the quantum field theory approach to gravity one would imagine that these entropies are related to ``sums over microstate geometries," rather than a single geometry.   

We believe that the most promising explanation for all of these results is the relation between geometry and hydrodynamics, exposed most clearly by Jacobson~\cite{Jacobson:1995ab} and subsequent works~\cite{Padmanabhan:2002ma,Padmanabhan:2003gd,Padmanabhan:2013nxa}.  Hydrodynamics is often invoked as a paradigm for effective field theory in pedagogic presentations of the renormalization group.  This is both correct and misleading.  If we have a large system with a non-degenerate ground state, the low energy long wavelength excitations of the system are usually described by variables parametrizing conserved currents.  These variables are treated as quantized fields with a cutoff and the physics is extracted by doing perturbation theory around a classical solution of the field equations representing either the ground state or the state created by some single high energy excitation.\footnote{Solitons or soft pion emission in baryon scattering are examples of non-ground state uses of effective field theory.}  This is the traditional realm of effective field theory in high energy physics, as well as quasi-particle physics in condensed matter theory.

On the other hand, when we study the non-equilibrium physics in a band of states with a very dense energy spectrum, we use the same current conservation equations as hydrodynamics, even far above the energy scale where the effective field theory of the previous paragraph loses validity.  A very similar dichotomous use of Einstein's equations is familiar to practitioners of AdS/CFT.  We use the (super)-gravity Lagrangian as a quantized field theory to compute Witten diagrams for correlators of small numbers of CFT operators in the unique ground state of the CFT.  On the other hand, we can understand the hydrodynamics of the strongly coupled CFT by solving the {\it classical} supergravity equations using the membrane paradigm on the stretched horizon of a black hole.  In this second context, it would be incorrect to calculate the quantum corrections to those equations and expect them to correctly describe the  fluctuation corrections to hydrodynamics.  The microstates on the black hole horizon are not well-described in terms of gravitons or other BPS particles in the bulk.

A recent derivation by Lucas and Banks~\cite{tbal} of hydrodynamic equations from the microscopic quantum mechanics of a large class of quantum lattice systems sheds some light on this issue.  It turns out that the hydrodynamic variables are the mutually commuting sub-Hamiltonians $H(X)$ of regions $X$ containing $V \gg 1$ lattice points.  The terms in the Hamiltonian coupling different regions are a small perturbation of $\sum_X H(X)$.  To leading order in this perturbation theory, the diagonal matrix elements of the density matrix in the basis of common eigenstates of all the $H(X)$ satisfy a Fokker-Planck equation, which is equivalent to a stochastic hydrodynamic equation for the time dependence of $E(X,t)$.   This is a classical statistical equation, showing no violation of Bayes' conditional probability rule.  It is expected that this is true to all orders in perturbation theory in powers of $1/V$.  The form of the equation will change but it will still obey Bayes' rule.   

The entropy $S(X)$ of the subsystem restricted to region $X$ appears explicitly in the Fokker-Planck equation, along with averaged squares of transition matrix elements between $H(X)$ eigenstates due to the terms in the full Hamiltonian that couple different regions.   The derivations in~\cite{tbal} are very general and apply to any system where that can be broken up into large subsystems with couplings between them that are small compared to each subsystem Hamiltonian.  In quasi-local theories, the hierarchy of Hamiltonians  $|| H(X) || \gg || H(X,Y) ||$ is a consequence of the scaling of surface versus volume in spatial geometries with non-negative curvature, but this is not the only way to obtain such a hierarchy. 

The point of this digression was to emphasize that hydrodynamic equations contain information about the entropies of large, weakly interacting subsystems, which is just the sort of information that has been extracted from Euclidean gravitational path integrals.  We do not yet have an elegant derivation of this connection, but we believe it will be an important part of the final understanding of the ``unreasonable effectiveness of Euclidean path integrals over metrics in the theory of quantum gravity."

Among the unfinished tasks along the lines of the current paper are an Euclidean rederivation of Verlinde's argument~\cite{Verlinde:2016toy} for the entropy deficit of causal diamonds containing black holes in Minkowski space, and an investigation of JT gravity, which would enable us to isolate degrees of freedom associated with finite causal diamonds.  We hope to return to these problems in future work.

~\\

{\bf Acknowledgements:}  PD acknowledges support from the National Science Foundation under Grant No. PHY-1719642 and from the US Department of Energy under Grant No. DE-SC0015655. The work of TB was partially supported by the Department of Energy under Grant No. DE-SC0010008.

\appendix
\label{ADMappx}
\section{Sum Rules from the ADM and Einstein-Hilbert Actions}

The total ADM action with all boundary terms included is equivalent to the sum of the Einstein-Hilbert action and all GHY terms. However, the separation of bulk and boundary contributions in the two calculations do not need to agree. Various bulk-boundary sum rules can be obtained by comparing the two actions. 

\subsection{ADM Action}
Here we follow the pedagogical discussion of~\cite{blaunotes} to summarize the boundary terms in the ADM action. The sign of the action used in~\cite{blaunotes} differs from the sign used here (Eq.~\ref{eq:action}), so we have adjusted the signs below accordingly.

Let the boundary include two spacelike hypersurfaces $\Sigma$, and in between them define a foliation $\Sigma_t$. Let $N^\alpha$ be a  vector field normal to the foliation satisfying $N^\alpha N_\alpha=\epsilon$. ($\epsilon=-(+)1$ for spacelike (timelike) $\Sigma_t$ respectively.) 
There can also be boundaries $B$ which intersect  the  hypersurface foliation on 2D boundaries $S_t$. There are two new boundary terms in the second-order ADM action associated with $B$: a Gauss-Codazzi boundary term  that is not a GHY term, and a GHY term. 

The terms associated with $B$ combine in such a way that they can be written as a boundary term on $S_t$ integrated over time. In the Hamiltonian there is a third boundary term involving the shift, which would also contribute to the first order form of the action. Including all boundary terms,
\begin{align}
I_{ADM} = -\frac{1}{16\pi}\int dt\left[\int_{\Sigma_t} \sqrt{h} N(\bar R +K^{ab}K_{ab}-K^2)+2\int_{S_t} d^2x \sqrt{s} N k_S\right]\nonumber\\
H_{ADM}=-\frac{1}{16\pi}\int_{\Sigma_t} d^3x\left(N{\cal H}+{\cal N}^a{\cal H}_a\right)+\frac{1}{8\pi} \int_{S_t} \sqrt{s}d^2x N k_S- \frac{1}{8\pi}\int_{S_t} d^2x {\cal N}_a \pi^{ab}r_b.
\end{align}
$k_S=s^{ab}\nabla_ar_b$ is the extrinsic curvature on ${S_t}$ and $s_{ab}=g_{ab}+N_aN_b-r_ar_b$ is the induced metric on ${S_t}$. We have also assumed $B$ and $\Sigma_t$ are orthogonal, $N_a r^a=0$.  $I_{ADM}$ is the same as the action computed with covariant methods,
\begin{align}
I_{ADM} = I_{EH}+I_{GHY,\Sigma} + I_{GHY,B}.
\label{eq:SADM}
\end{align}
On the constraint surface, 
\begin{align}
I_{ADM}&\rightarrow  \int dt d^3x\,\pi^{ab}\dot h_{ab}- \frac{1}{8\pi}\int_{S_t} d^2x \sqrt{s} N k_S,\nonumber\\
H_{ADM}&\rightarrow \frac{1}{8\pi} \int_{S_t} d^2x \left(\sqrt{s} N k_S-  {\cal N}_a \pi^{ab}r_b\right).
\end{align}
So  we have
\begin{align}
I_{ADM} = \int dt \left[\left(\int_{\Sigma_t} d^3x\,\pi^{ab}\dot h_{ab}\right)-{ H}_{ADM} -\frac{1}{8\pi}\int_{S_t} d^2x {\cal N}_a\pi^{ab}r_b\right]
\end{align}
where $H_{ADM}$ is just the boundary terms in the preceding equation summed over all boundaries $S_t$.
  In many cases of interest  the last term in $I_{ADM}$ vanishes because the shift is zero, and the first term vanishes because $\dot h_{ab}=0$.  So $I_{ADM}=-\int dt H_{ADM} = -\frac{1}{8\pi}\int dt \int_{S_t}d^2x\sqrt{s}Nk_S$ in these circumstances.
 
 Here we have left off regulators. In Schwarzschild $H_{ADM}$ at radial infinity is divergent. It can be cancelled by a subtraction similar to what is done for GHY terms.

Most of the preceding discussion carries over to Euclidean signature. However, in cases where the continuation of the timelike Killing vector has a fixed point, it is especially convenient to define radial time slices on the Euclidean manifold. Then we have to contend with the fact that the time slices intersect at the point. To handle this case, we use the method of Refs.~\cite{Banados:1993qp,Teitelboim:2002cv}. 
We excise a small disc of radius $\epsilon$ around the point of convergence and use the Lagrangian action inside the disc, retaining the ADM action outside. Since the two actions are equal, this surgery makes no difference to the total result. The procedure introduces two boundary terms on the disc: an outward-pointing GHY term, and an inward-pointing ADM boundary term.

\subsection{Schwarzschild}
Now we compute the ADM action of Euclidean Schwarzschild, taking $\beta=1/(8\pi M)$ to remove the conical singularity at $2M$. Our foliation is adapted to the static coordinates, so that  the normal to the time slices is $N^a\propto \partial_t$. The lapse is $N=\sqrt{f}=\sqrt{1-2M/r}$,  the shifts are zero, and the induced metric on the time slices is $h_{ab}dx^adx^b= f^{-1/2}dr^2+r^2d\Omega^2$. 

The time slices intersect at $r=2M$, so we will use the excision technique there. Then there are two ADM boundary terms, one at infinity pointing outward, and one on the disc pointing inward. There are no ADM bulk terms. There is a GHY term on the disc pointing outward, and no Einstein-Hilbert bulk term inside the disc. 

The ADM boundary terms are of the form $-\int H dt$, with the Hamiltonian boundary terms evaluated on $S_t$'s that are 2-spheres with the induced metric $s_{ab}dx^adx^b = r^2d\Omega^2$. The unit normal to the $S_t$ surfaces that is orthogonal to $N^a$ is $r^a=\pm\sqrt{f}\partial_r$.  We obtain
\begin{align}
k_S&=\pm\frac{1}{\sqrt{h}}\partial_r\left(\sqrt{h} \sqrt{f}\right)=\pm\frac{2}{r}\sqrt{f}\nonumber\\
\Delta I_E &= -\frac{1}{8\pi}\int H dt = \left(-\frac{1}{8\pi}\right)(4\pi)(\beta)\left[-r^2 N |k_S|\bigg|_{r\rightarrow 2M^+}+r^2 N |k_S|\bigg|_{r\rightarrow\infty^-}\right]\nonumber\\
&=-\beta\left[rf\bigg|_{r\rightarrow 2M^+}+rf\bigg|_{r\rightarrow\infty^-}\right]\nonumber\\
&=-\beta\left[0+(r-M)\bigg|_{r\rightarrow\infty}\right]\nonumber\\
&\rightarrow \beta M.
\end{align}
In the last line we have subtracted the same boundary term from flat space to regularize the divergence. In this case it amounts to subtracting the same computation with $M=0$. We also see that the Hamiltonian boundary term at $2M$ vanishes since it is proportional to $f$.

The GHY term on the disc is the same as one we already computed for the Schwarzschild diamond, but with opposite sign because the normal points to larger $r$ on the disc. We get
\begin{align}
\sqrt{h} K\bigg|_{r\rightarrow 2M} &=\frac{\sqrt{h}}{\sqrt{g}}\partial_r\left( \sqrt{g}\sqrt{f}\right)|_{r=2M}\nonumber\\
&= (2r-3M)_{r=2M}\sin(\theta)\nonumber\\
\Delta I_E &=-\frac{1}{8\pi}(4\pi)(\beta)(M)\nonumber\\
&=-\frac{\beta M}{2} =-\frac{A_{BH}}{4}.
\end{align}

Adding the pieces up, we get
\begin{align}
I_E = \frac{\beta M}{2} = \frac{A_{BH}}{4}
\end{align}
which is the expected value in the canonical ensemble, $I_E = \beta F = \beta M- S =  A/4$.

One advantage of the ADM formalism is it makes it clear how to transition to the microcanonical ensemble of the horizon degrees of freedom: we just drop the boundary terms at infinity, which contributed $\Delta I_E=+\beta M$. Then we find 
\begin{align}
\log(Z)_{micro} = +S = +A_{BH}/4
\end{align}
computed entirely by the outward-pointing GHY term in the infinitesimal disc around $r=2M$~\cite{Banados:1993qp}.

\subsection{dS}
Euclidean dS is topologically a sphere, with no boundaries.  In the usual computation, the free energy is saturated by the Einstein-Hilbert term. However, it can also be obtained purely from a boundary term, as can be seen from the ADM form of the action.

We take static patch coordinates. The time coordinate is periodic, with period set to $\beta=2\pi L$ to remove the conical singularity at $r=L$. 
Again we take our foliation to be constant-$t$ slices, so that  the normal to the spatial slices is $N^a\propto \partial_t$, the lapse is $N=\sqrt{f}=\sqrt{1-r^2/L^2}$,  the shifts are zero, and the induced metric on the spatial slices is $h_{ab}dx^adx^b= f^{-1/2}dr^2+r^2d\Omega^2$. 

As in the Schwarzschild case, the slices intersect at at point, in this case $r\rightarrow L$. Employing the excision technique, there is one ADM boundary term on the disc pointing outward toward  $r=L$, and no ADM bulk terms. There is a GHY term on the disc pointing away from the disc toward smaller $L$, and the Einstein-Hilbert bulk term inside the disc in the limit that its size goes to zero. 

The ADM boundary terms are of the form $-\int H dt$, with the Hamiltonian boundary terms evaluated on $S_t$'s that are 2-spheres with the induced metric $s_{ab}dx^adx^b = r^2d\Omega^2$. The unit normal to the $S_t$ surfaces that is orthogonal to $N^a$ is $r^a=\sqrt{f}\partial_r$.  We get
\begin{align}
k_S&=\frac{1}{\sqrt{h}}\partial_r\left(\sqrt{h} \sqrt{f}\right)=\frac{2}{r}\sqrt{f}\nonumber\\
\Delta I_E &= -\frac{1}{8\pi}\int H dt = \left(-\frac{1}{8\pi}\right)(4\pi)(\beta)\left(r^2 N |k_S|\bigg|_{r\rightarrow L}\right)\nonumber\\
&=0.
\end{align}
The Hamiltonian boundary term at $L$ vanishes since it is proportional to $f$.

The GHY term on the disc can be obtained from the maximal causal diamond computation in Sec.~\ref{sec:dS}. We get
\begin{align}
\Delta I_E = -\frac{A_{dS}}{4}.
\end{align}

Adding the pieces up, 
\begin{align}
I_E = 0-\frac{A_{dS}}{4} = -\frac{A_{dS}}{4}
\end{align}
which is the expected value in the canonical ensemble, $I_E = \beta F  = -S = -A/4$.

We see that the equivalence between the ADM and Lagrangian forms of the action results in a sum rule for Euclidean dS,
\begin{align}
I_{GHY,r=L}=I_{EH}.
\end{align}

\subsection{SdS}
Euclidean SdS is similar to dS, but it has at least one conical singularity, and we have to delete that point. The deleted point is treated differently in the Lagrangian and ADM analyses of the action. In a Lagrangian analysis, we put a GHY boundary around the deleted point, and it will be nonzero. In an ADM analysis, we can choose the time slicing so that the deleted point is  one of  two places where the time slices intersect. Ref.~\cite{Teitelboim:2002cv} generalized the prescription of~\cite{Banados:1993qp} to this case: we are instructed to draw a disc around the conical singularity,  delete the whole interior, and finally  take the disc radius to zero. The result of this is we have no action contribution from the  interior of the disc, bulk or boundary. We have only in the inward-pointing Hamiltonian boundary term from the exterior of the disc. 

At the other point in SdS where the ADM time slices intersect, the manifold is smooth. There we have to do the same analysis as in the previous sections.

We already did the Lagrangian analysis above for the case $\beta=1/T_c$, and the case $\beta=1/T_b$ is similar. 
 If we choose $\beta= 1/T_{c(b)}$, the total action is $I_E = -\pi r_{c(b)}^2$. The extrinsic curvature on surfaces at generic constant $r$ (that we will also need for the ADM analysis in the prescription of~\cite{Teitelboim:2002cv}) is
 \begin{align}
 &\sqrt{h}K = -\left(3r^3/L^2 +3M-2r\right)\sin\theta
 \end{align}
 with normal pointing to larger $r$. 
 
In the ADM analysis, the bulk term  is always zero since we have static coordinates and we satisfy the constraints. Furthermore, the Einstein-Hilbert bulk terms inside the discs go to zero smoothly in the limit that the disc radius goes to zero. Thus we have only boundary terms to evaluate: two ADM boundary terms at $r_{b,c}$ and one GHY boundary term from the disc around whichever of these two points \emph{isn't} deleted. The ADM boundary terms vanish for the same reason they do in the Schwarzschild case at $r=2M$: $\int H dt$ is proportional to $f=1-2M/r-r^2/L^2$, which vanishes at both $r_b$ and $r_c$. All that is left is the outward-pointing GHY boundary term from the disc located at whichever horizon is smooth. We have
\begin{align}
I_E &= -\frac{1}{8\pi}(4\pi)(1/T_b) \left(-(3r_b^3/L^2 +3M-2r_b)\right)\nonumber\\
&=-\pi r_b^2~~~~~~~~~~(T=T_b)\nonumber\\ 
I_E &= -\frac{1}{8\pi}(4\pi)(1/T_c) \left(+(3r_c^3/L^2 +3M-2r_c)\right)\nonumber\\
&=-\pi r_c^2~~~~~~~~~~(T=T_c) .\nonumber\\ 
\end{align}
matching exactly the Lagrangian analysis.

The ADM analysis explains why putting outward pointing GHY boundaries at $r_b$ and $r_c$ and adding to the bulk Einstein-Hilbert action gives zero. The ADM and Lagrangian actions amount to the sum rule
\begin{align}
I_{EH}+I_{GHY,r_b}+I_{GHY,r_c} = 0.
\end{align}

These results show that $I_E$ is computing $-S_{horizon}$ for both choices of $T$. $F/T_c=-S_c$ anyways for the cosmological horizon, but for the black hole, it only makes sense if $-I_E$ is computing the log of the partition function for the microcanonical ensemble. The ADM calculation  clarifies why this is so.  In the large $L$ limit, the ADM analysis with $T=T_b$ maps onto the \emph{microcanonical} computation in the asymptotically flat Schwarzschild case, where we remove $\beta M$ (i.e., we drop the ADM boundary term at infinity.) 

\bibliography{diamond_refs}{}
\bibliographystyle{utphys}

\end{document}